\documentclass[]{pasj02}

\usepackage{amsmath}
\usepackage{array}
\usepackage{tabularx}
\usepackage{natbib}
\usepackage[switch,mathlines]{lineno}

\let\leftrightarrows\relax
\let\gtrsim\relax

\let\lesssim\relax

\let\varnothing\relax

\usepackage{amssymb}

\Received{$\langle$reception date$\rangle$}
\Accepted{$\langle$acception date$\rangle$}
\Published{$\langle$publication date$\rangle$}

\begin{document}

\title{Grain growth in protoplanetary disks in the Upper Scorpius revealed by millimeter-wave spectral indices}

\author{Tau Bito\altaffilmark{1}}
\altaffiltext{1}{Department of Astronomy, Kyoto University, Kitashirakawa-Oiwake-cho, Sakyo-ku, Kyoto 606-8502, Japan}
\author{Akimasa Kataoka\altaffilmark{2}}
\altaffiltext{2}{National Astronomical Observatory of Japan, 
 2-21-1 Osawa, Mitaka, Tokyo 181-8588, Japan }
\author{Takahiro Ueda\altaffilmark{3}}
\altaffiltext{3}{Center for Astrophysics \textbar{} Harvard \& Smithsonian, 60 Garden Street, Cambridge, MA 02138, USA}
\author{Luca Ricci\altaffilmark{4}}
\altaffiltext{4}{Department of Physics \& Astronomy, California State University Northridge, 18111 Nordhoff Street, Northridge, CA 91330, USA}
\author{Tilman Birnstiel\altaffilmark{5,6}}
\altaffiltext{5}{University Observatory, Faculty of Physics, Ludwig-Maximilians-Universität München, Scheinerstr. 1, 81679 Munich, Germany}
\altaffiltext{6}{Exzellenzcluster ORIGINS, Boltzmannstr. 2, 85748 Garching, Germany}
\author{John Carpenter\altaffilmark{7}}
\altaffiltext{7}{Joint ALMA Observatory, Avenida Alonso de Córdova 3107, Vitacura, Santiago, Chile}

\KeyWords{protoplanetary disks --- planets and satellites: formation --- radio continuum: planetary systems}

\maketitle

\begin{abstract}
The measurement of dust size from millimeter-wavelength spectra provides direct constraints on grain growth in protoplanetary disks. 
The spectral indices between $0.88~\mathrm{mm}$ and $2.9~\mathrm{mm}$ have been measured in multiple young star-forming regions, such as Taurus, Ophiuchus, and Lupus, which have ages of 1-3 Myr. 
These spectral indices are as low as 2-3, suggesting that grains in disks are much larger than those in the interstellar medium.
In this study, we analyze the ALMA archival data of 23 disks in the Upper Scorpius region. 
The observed wavelength is $2.9~\mathrm{mm}$ in Band 3, the angular resolution is 3$\farcs$3 $\times$ 2$\farcs$1, which is not high enough to resolve the targets, and the rms noise is below $0.075~\mathrm{mJy}~\mathrm{beam}^{-1}$ for almost all sources.
Together with the literature values of the Band 7 fluxes of the same targets, we find that the average spectral index of the disks in the Upper Scorpius region is $\alpha_\mathrm{mm}=2.09\pm0.10$, which is equal to or slightly smaller than those at the other younger regions.
To explain the relationship between the fluxes and spectral indices of the disks in the Taurus, Ophiuchus, Lupus, and Upper Scorpius regions, we construct simple disk evolution models.
The observations are best reproduced by models in which the inner radius of the disk increases.
This suggests that a substantial amount of dust mass must persist in the outer disk regions where the dust temperature is lower than 20 K even at late evolutionary stages.
These findings offer key insights into the grain growth and the temporal evolution of protoplanetary disks.
\end{abstract}

\section{Introduction}
Planet formation begins with the growth of micron-sized dust particles within protoplanetary disks, eventually leading to the formation of planetesimals, terrestrial planets, and the cores of gas giants. Dust grains in these disks grow by sticking together, fragmenting into smaller ones or drifting radially, depending on the disk properties \citep{Birnstiel2024}. 

The grain size in disks is observationally deduced from the millimeter spectral index of the dust continuum emission.
If the disk is optically thin and the radiation is emitted in the Rayleigh–Jeans regime, the relationship $\alpha \simeq \beta +2$ holds, where $\alpha$ is the spectral index of the flux density ($F_\nu \propto \nu^\alpha$), and $\beta$ is the spectral index of the mass absorption dust opacity coefficient ($\kappa_\nu \propto \nu^\beta$).
Under this assumption, $\beta$ depends on the maximum grain size $a_\mathrm{{max}}$ and overall grain size distribution \citep{Natta2007}.
The typical value of $\beta$ is 1.7 for interstellar medium (ISM; \citealp{Finkbeiner1999}).
As the grain size increases from micron sizes, $\beta$ initially stays nearly constant, then rises at the opacity cliff around $a_\mathrm{max}=\lambda/(2\pi)$, and subsequently starts to decrease (e.g., \citealp{Beckwith1991,Miyake1993,Draine2006,Ricci2010Taurus,Birnstiel2018}).
Multi-wavelength observations in regions like Lupus, Taurus, and Ophiuchus have detected spectral indices lower than those of ISM, supporting the theory of grain growth within these disks \citep{Tazzari2021,Ricci2010Taurus,Ricci2010Ophiuchus}. 

However, the validity of this discussion diminishes in regions with optically thick areas \citep{Ricci2012} or thermal emission deviating from the Rayleigh-Jeans regime.
Advances in high-resolution imaging and spectral analysis, especially with the Atacama Large Millimeter/Submillimeter Array (ALMA), have enabled spatially resolved studies of disk structures \citep{Perez2012,Tsukagoshi2016,Andrews2018,Macias2021}. 
These observations have uncovered intricate substructures, such as rings and gaps, which may influence spectral interpretations and affect our understanding of dust evolution \citep{Andrews2020}.
In particular, the localized dust distribution and/or the compactness of disks make them more optically thick than they appear, highlighting the need for observations at longer wavelengths where the disk is more optically thin \citep{CG19,Ueda2022,Guidi2022}.

Photometric surveys are essential for investigating the temporal evolution of dust in disks in a statistical way.
The temporal evolution of dust within disks has been studied by comparing individual objects at different evolutionary phases (Class 0, I, II) within the same region \citep{Lada1984,Andre1993,Tobin2020} or through global comparisons on a region-by-region basis \citep{Ansdell2016,Barenfeld2016,Testi2022}.
Upper Scorpius is a nearby star-forming region with an estimated age of around 5–10 Myr \citep{Pecaut2016}, significantly older than the other regions such as Taurus and Lupus ($\sim$1-3 Myr; \citealp{Alcala2017}), offering crucial insights into the temporal evolution of dust in disks.

In this study, we analyze ALMA Band 3 data at a wavelength of 2.9 mm for disks in the Upper Scorpius (USco) region and derive spectral indices by comparing it with Band 7 data.  
We then examine how these spectral indices differ from those in other regions.
We also make a simple model to consider the effects of optically thick and cooler regions.
The goal of this paper is to gain insight into dust growth in the disk by comparing observations by region and by comparing models with observations.

\section{Sample Selection and Observation}
In this study, we used ALMA Band 3 observation data of 23 disk-bearing young stellar objects (YSOs) in the Upper Scorpius star-forming region.
These observations were obtained on June 20 and 22, 2016 (project ID: 2015.1.00819.S, PI: Ricci L.).
The observed disks correspond to the brightest objects in the Band 7 survey \citep{Barenfeld2016} with the exception of 2MASS J1653456-2242421.
The derived stellar properties are given in Table \ref{sample_table}.
Stellar spectral types and disk types were taken from \citet{LuhmanMamajek2012} and \citet{Luhman2020}, respectively.
Distances were calculated using the parallaxes of \citet{GaiaDR3}, except for J16141107-2305362, J16014086-2258103, J16135434-2320342 and J16062196-1928445.
For these objects, the average distance of the other objects ($141.6~\mathrm{pc}$) was applied.

\begin{longtable}{*{6}{c}}
\caption{Stellar properties.}
\label{sample_table}
\hline
Source&SpT&Disk Type&R.A.&Dec.&Distance\\
(2MASS)&&&(ICRS)&(ICRS)&(pc)\\
\hline
\endhead
\hline
\multicolumn{6}{p{0.8\textwidth}}{\textbf{Notes:} 
Stellar spectral types and disk types were taken from \citet{LuhmanMamajek2012} following \citet{Barenfeld2016}. We use the definitions of the evolutionary stages of disks adopted by \citet{Espaillat2012}.
Distances are computed using Gaia DR3 measurements \citep{GaiaDR3}.
Values marked with * have been averaged from other sources as the data for Gaia DR3 were not available.
}\\
\endlastfoot
J16113134-1838259&K5&Full&16:11:31.34&-18.38.25.9&$132.1{\pm1.2}$\\
J16042165-2130284&K2&Transitional&16:04:21.65&-21.30.28.4&$145.3{\pm0.6}$\\
J15583692-2257153&G7&Full&15:58:36.92&-21.30.28.4&$167.2{\pm0.6}$\\
J16090075-1908526&K9&Full&16:09:00.75&-19.08.52.6&$137.4{\pm0.4}$\\
J16082324-1930009&K9&Full&16:08:23.24&-19.30.00.9&$137.7{\pm0.4}$\\
J16142029-1906481&M0&Full&16:14:20.29&-19.06.48.1&$139.9{\pm1.0}$\\
J16154416-1921171&K5&Full&16:15:44.16&-19.21.17.1&$126.7{\pm1.3}$\\
J16075796-2040087&M1&Full&16:07:57.96&-20.40.08.7&$136.8^{+2.5}_{-2.4}$\\
J16072625-2432079&M3.5&Full&16:07:26.25&-24.32.07.9&$144.5{\pm0.6}$\\
J16024152-2138245&M4.75&Full&16:02:41.52&-21.38.24.5&$140.4{\pm1.2}$\\
J16054540-2023088&M2&Full&16:05:45.40&-20.23.08.8&$138.7{\pm0.8}$\\
J16135434-2320342&M4.5&Full&16:13:54.34&-23.20.34.2&${}*141.6{\pm1.8}$\\
J16123916-1859284&M0.5&Full&16:12:39.16&-18.59.28.4&$135.5{\pm0.4}$\\
J15582981-2310077&M3&Full&15:58:29.81&-23.10.07.7&$141.0{\pm1.0}$\\
J15530132-2114135&M4&Full&15:53:01.32&-21.14.13.5&$143.1{\pm1.0}$\\
J16020757-2257467&M2.5&Full&16:02:07.57&-22.57.46.7&$140.6{\pm0.4}$\\
J16111330-2019029&M3&Full&16:11:13.30&-20.19.02.9&$153.8{\pm0.7}$\\
J16141107-2305362&K2&Full&16:14:11.07&-23.05.36.2&${}*141.6{\pm1.8}$\\
J16181904-2028479&M4.75&Evolved&16:18:19.04&-20.28.47.9&$139.7{\pm1.2}$\\
J16035767-2031055&K5&Full&16:03:57.67&-20.31.05.5&$142.9{\pm0.4}$\\
J16062196-1928445&M0&Transitional&16:06:21.96&-19.28.44.5&${}*141.6{\pm1.8}$\\
J16001844-2230114&M4.5&Full&16:00:18.44&-22.30.11.4&$162.3^{+11.6}_{-10.1}$\\
J16014086-2258103&M4&Full&16:01:40.86&-22.58.10.3&${}*141.6{\pm1.8}$\\
\end{longtable}

The continuum spectral windows of the observations were centered on 97.995, 99.932, 109.995, and 111.995 GHz with band widths of 1.875 GHz and channel widths of 15.625 MHz, respectively.
The array configuration used forty-five 12 m antennas for the 11 faint samples (with Band 3 fluxes less than approximately $0.9~\mathrm{mJy}$) and forty-one 12 m antennas for the 12 bright samples (with Band 3 fluxes greater than approximately $0.9~\mathrm{mJy}$) with maximum baselines of 331.0 m.
The integration times are 9 min per source on the faint targets and 3 min per source on the bright targets, achieving an rms noise of 28 and 67~$\mathrm{\mu Jy/beam}$, respectively.
Data calibration and imaging were performed using CASA 4.5.2 and 4.5.3 \citep{McMullin2007}.
We extracted continuum images from the calibrated visibilities by averaging over the continuum channels.
The average continuum beam size, adopting natural weighting, was 3.3$\times$2.1 arcsec${}^2$ ($\approx$ 460$\times$280 au${}^2$ at the distance of these sources).

We measured continuum flux densities by fitting point-source to the visibility data with \texttt{uvmodelfit} in CASA.  
We confirmed that all objects were unresolved in continuum images produced by the CASA tclean task.
The point-source model has three free parameters: integrated flux density ($F_\mathrm{cont}$), right ascension offset from the phase center ($D_a$), and declination offset from the phase center ($D_d$).  The initial parameters for the fit were the same for all objects: $F_\mathrm{cont}=0.005~\mathrm{Jy}$, $D_a=0.0$, $D_d=0.0$.

\section{Results}
\begin{longtable}{*{6}{c}}
\caption{Continuum flux measurements and $0.88-2.9~\mathrm{mm}$ spectral indices.}
\label{result_table}
\hline
Source & Beam Size  & rms & $F_{2.9\mathrm{mm}}$ & $F_{0.88\mathrm{mm}}$ & $\alpha_\mathrm{mm}$\\
&($\mathrm{arcsec \times arcsec}$)&($\mathrm{mJy~beam^{-1}}$)&($\mathrm{mJy}$)&($\mathrm{mJy}$)&\\
\hline
\endhead
\hline
\multicolumn{6}{l}{\textbf{Note:}
$F_{0.88\mathrm{mm}}$ are reported by \citet{Barenfeld2016}. 
}\\
\endlastfoot
J16113134-1838259&3.31$\times$2.08&0.075&$51.824^{\pm0.021}$&$903.56\pm0.85$&$2.43\pm0.09$\\
J16042165-2130284&3.27$\times$2.08&0.038&$5.122^{\pm0.021}$&$218.76\pm0.81$&$3.19\pm0.10$\\
J15583692-2257153&3.25$\times$2.08&0.042&$5.033^{\pm0.021}$&$174.92\pm0.27$&$3.01\pm0.10$\\
J16090075-1908526&3.29$\times$2.07&0.041&$3.313^{\pm0.022}$&$47.28\pm0.91$&$2.26\pm0.10$\\
J16082324-1930009&3.28$\times$2.07&0.042&$3.147^{\pm0.022}$&$43.19\pm0.81$&$2.22\pm0.10$\\
J16142029-1906481&3.30$\times$2.07&0.047&$3.193^{\pm0.021}$&$40.69\pm0.22$&$2.16\pm0.10$\\
J16154416-1921171&3.31$\times$2.08&0.046&$2.068^{\pm0.021}$&$23.57\pm0.16$&$2.07\pm0.10$\\
J16075796-2040087&3.27$\times$2.07&0.300&$2.643^{\pm0.021}$&$23.49\pm0.12$&$1.86\pm0.10$\\
J16072625-2432079&3.24$\times$2.08&0.047&$1.054^{\pm0.021}$&$13.12\pm0.24$&$2.14\pm0.10$\\
J16024152-2138245&3.24$\times$2.07&0.042&$1.109^{\pm0.021}$&$10.25\pm0.19$&$1.89\pm0.10$\\
J16054540-2023088&3.27$\times$2.07&0.045&$0.991^{\pm0.022}$&$7.64\pm0.15$&$1.73\pm0.10$\\
J16135434-2320342&3.27$\times$2.08&0.038&$1.202^{\pm0.021}$&$7.53\pm0.13$&$1.56\pm0.10$\\
J16123916-1859284&3.22$\times$2.14&0.037&$0.467^{\pm0.015}$&$6.01\pm0.29$&$2.17\pm0.11$\\
J15582981-2310077&3.17$\times$2.14&0.027&$0.651^{\pm0.015}$&$5.86\pm0.18$&$1.87\pm0.10$\\
J15530132-2114135&3.17$\times$2.14&0.032&$0.681^{\pm0.015}$&$5.78\pm0.14$&$1.82\pm0.10$\\
J16020757-2257467&3.19$\times$2.14&0.030&$0.219^{\pm0.015}$&$5.26\pm0.27$&$2.70\pm0.12$\\
J16111330-2019029&3.22$\times$2.14&0.026&$0.758^{\pm0.015}$&$4.88\pm0.16$&$1.58\pm0.10$\\
J16141107-2305362&3.23$\times$2.15&0.029&$0.705^{\pm0.015}$&$4.77\pm0.14$&$1.62\pm0.10$\\
J16181904-2028479&3.23$\times$2.14&0.021&$0.521^{\pm0.015}$&$4.62\pm0.12$&$1.85\pm0.10$\\
J16035767-2031055&3.22$\times$2.14&0.033&$0.453^{\pm0.015}$&$4.30\pm0.39$&$1.91\pm0.13$\\
J16062196-1928445&3.21$\times$2.14&0.029&$0.616^{\pm0.015}$&$4.08\pm0.52$&$1.61\pm0.15$\\
J16001844-2230114&3.18$\times$2.14&0.025&$0.359^{\pm0.015}$&$3.89\pm0.15$&$2.02\pm0.11$\\
J16014086-2258103&3.22$\times$2.14&0.023&$0.335^{\pm0.015}$&$3.45\pm0.14$&$1.98\pm0.11$\\
\end{longtable}

\subsection{Flux at $2.9~\mathrm{mm}$ and spectral indices}
Table \ref{result_table} summarizes our results for the $2.9~\mathrm{mm}$ continuum flux densities and disk-integrated spectral indices. 
Note that the uncertainties of the fluxes are statistical errors and do not include the absolute flux calibration error, which is estimated to be 5\% for Band 3 data according to the ALMA technical handbook.
No emission lines were detected.

The flux at $2.9~\mathrm{mm}$ from the brightest disk is $51.824\pm0.021~\mathrm{mJy}$, which is 690 times the noise.
The faintest one is $0.335~\mathrm{mJy}$, which is 20 times the noise.
Spectral indices $\alpha_\mathrm{mm}$ were computed using the $0.88~\mathrm{mm}$ fluxes measured by \citet{Barenfeld2016} and assuming a simple power law between $0.88$ and $2.9~\mathrm{mm}$ : $F_\nu \propto \nu^{\alpha_\mathrm{mm}}$.
Their errors include the 10\% flux calibration error.
The mean spectral index is $\alpha_\mathrm{mm} = 2.09\pm0.10$.

\subsection{Comparison to other regions}
\begin{figure}
    \includegraphics[width=80mm]{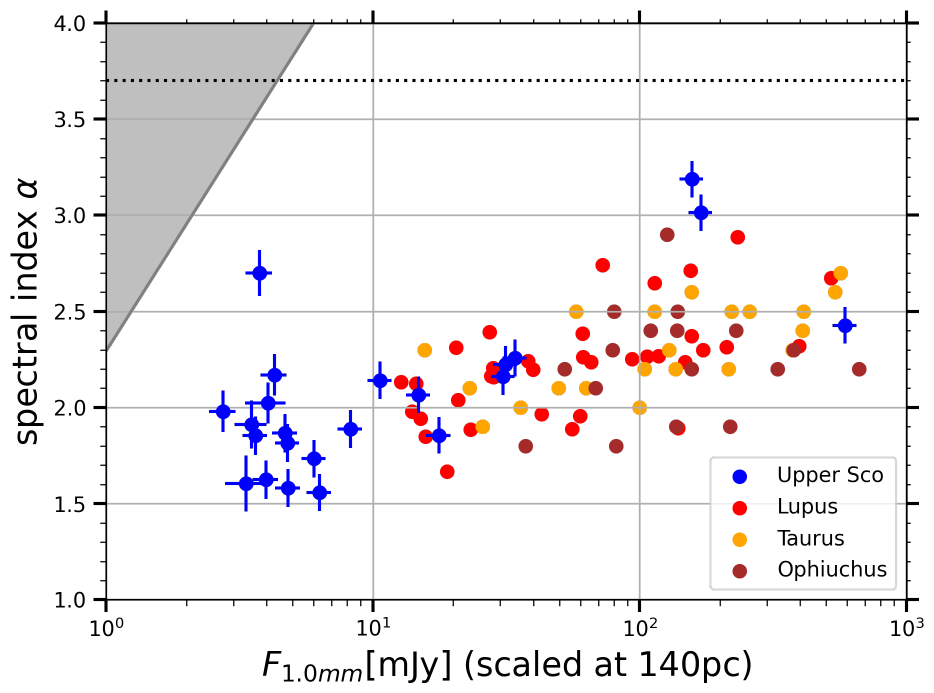}
    \caption{Spectral indices between $0.88$ and $2.9~\mathrm{mm}$ as a function of integrated $1~\mathrm{mm}$ flux for USco disks (blue), Lupus (red; \citealp{Tazzari2021}) , Taurus (orange; \citealp{Ricci2010Taurus}) and Ophiuchus (brown; \citealp{Ricci2010Ophiuchus}).  
    The uncertainties in the flux measurements, shown as blue horizontal bars in the figure, are due to a 10\% calibration uncertainty.
    The errors on the spectral indices, shown as blue vertical bars, are propagated from the flux uncertainties at the two wavelengths.
    The dark shaded region represents the sensitivity cut-off of our ALMA observations which is caused by the difference between the fluxes at $0.88$ and $2.9~\mathrm{mm}$.
    %{Alt text: Scatter plot.}
    }
    \label{spectral_index}
\end{figure}

To compare the spectral indices between star-forming regions, figure \ref{spectral_index} shows the spectral indices as a function of the $1~\mathrm{mm}$ flux of the brightest disks in the regions of Lupus, Taurus, Ophiuchus, and Upper Scorpius.
The $1~\mathrm{mm}$ fluxes were calculated under the assumption of constant spectral indices between the $2.9~\mathrm{mm}$ and $0.88~\mathrm{mm}$ data.
The spectral indices of disks in Lupus are taken from \citet{Tazzari2021} and those in Taurus and Ophiuchus are taken from a compilation by \citet{Ricci2010Ophiuchus} and \citet{Ricci2010Taurus}, where the scaled flux densities are recalculated with the corrected distances (see Appendix).

The fluxes of disks in USco are significantly lower than those in the other regions, corresponding to lower disk masses.
The mean spectral index of disks in Upper Scorpius is $\alpha_\mathrm{mm} = 2.09\pm0.10$, while those in Lupus, Taurus and Ophiuchus are 2.2, 2.3 and 2.2 respectively.
There are many more disks with spectral indices below 2 in the USco region than in other regions.
The overall trend is that the smaller the flux, the smaller the spectral index.
Note, however, that J16020757-2257467 has a relatively small $F_{1\mathrm{mm}}$ but a large $\alpha$ of 2.7, and thus deviates from this trend.

\section{Discussion}
The observed millimeter spectral indices of protoplanetary disks provide a key diagnostic of the physical properties of the dust they contain. 
The evolution of dust disks is governed by a range of complex processes such as sticking growth, radial drift, viscous spreading with gas \citep{Lyden1974}, gap formation by protoplanets (e.g., \citealp{Dipierro2015}) and magnetic field effects \citep{Flock2015}.
A comprehensive treatment of these mechanisms is beyond the scope of this paper. 
Instead, we construct a set of parametric disk models to explore how different modes of mass loss affect millimeter observables. 
These models are not designed to reproduce specific physical processes of disk dispersal, but rather to identify the necessary dust grain properties that are consistent with the observed millimeter emission from dust disks. 
In this section, we first describe the setup of our simple models and then discuss the resulting implications.

\subsection{Simple disk models}
Excluding variations in dust composition or grain size, the structure of a simple dust disk can be characterized by three free parameters: the inner radius ($R_\mathrm{min}$), the outer radius ($R_\mathrm{max}$), and the normalization of the surface density ($\Sigma_0$).
Now, the surface density distribution is given by
\begin{align}
\Sigma(r) = 
\begin{cases} 
  \Sigma_0 \left(\frac{r}{1\,\mathrm{au}}\right)^{-1} & (R_{\mathrm{min}} < r < R_{\mathrm{max}}) \\
  0 & (\text{otherwise}).
\end{cases}
\end{align}
We adopt a $r^{-1}$ surface density profile, which is shallower than that of the minimum-mass solar nebula model \citep{Hayashi1981} but still consistent with recent observational constraints suggesting moderately declining profiles \citep{Law2021}.

We consider four scenarios (Cases A–D), each varying one parameter to represent a different mode of disk mass loss, while holding the others fixed. 
In Case A, the outer disk radius decreases from $1000$ au to $1$ au, while the inner radius and the surface density normalization are kept constant at $R_\mathrm{min}=0.1$ au and $\Sigma_0=4.23~\mathrm{g\,cm^{-2}}$. 
In Case B, the inner disk radius increases from $0.1$ au to $100$ au, with $R_\mathrm{max}=100$ au and $\Sigma_0=4.23~\mathrm{g\,cm^{-2}}$ fixed. 
In Case C, the surface density normalization decreases from $42.3$ to $0.0423~\mathrm{g\,cm^{-2}}$, while the inner and outer radii remain at $0.1$ au and $100$ au, respectively. 
Case D follows the same setup as Case C, in which the surface density normalization is reduced while keeping $R_\mathrm{min}=0.1$ au and $R_\mathrm{max}=100$ au fixed, but with additional gaps imposed in the disk. 
Specifically, $\Sigma(r)$ is set to zero for $20~\mathrm{au}<r<55~\mathrm{au}$, $75~\mathrm{au}<r<90~\mathrm{au}$, mimicking the gap structure observed in HD 163296 \citep{Doi2023}. 
The surface density profile is then renormalized such that the initial total dust mass is $1000\,M_\oplus$, and subsequently reduced until the mass reaches $1\,M_\earth$.

According to \citet{Testi2022}, most of the stellar luminosities in these regions fall within the range of $0.01$ to $1L_{\odot}$.
Stars in USco tend to be fainter than those in other regions, but the mean luminosity of our sample is $0.25L_\odot$ because we are using only the brighter ones within USco.

Adopting the midplane temperature profile
\begin{align}
T_\mathrm{mid}(r) = \left(\frac{\phi L_{*}}{8\pi r^2 \sigma_\mathrm{SB}}\right)^{0.25}\quad[K]
\end{align}
as used in \citet{Huang2018}, and assuming a stellar luminosity of $L_{*} = 0.25L_{\odot}$ and a flaring angle of $\phi = 0.04$, we obtain the following expression:
\begin{align}
T_\mathrm{mid}(r) = 100\left(\frac{r}{1\mathrm{au}}\right)^{-0.5}\quad[K].
\end{align}
This temperature profile is used consistently across all four models, and is assumed to remain constant over time.

We adopted a dust opacity 
\begin{align}
    \kappa_{\nu}=2.7\left(\frac{\nu}{\nu_0}\right)^\beta~[\mathrm{cm^2~g^{-1}}],
\end{align}
while $\nu_0 = 341~\mathrm{GHz}$ ($0.88~\mathrm{mm}$), following \citet{Barenfeld2016}.
For each model, we test five values for the opacity index, $\beta = 0, 0.5, 1.0, 1.5, 2.0$ as an indicator of dust grain size \citep{Ricci2010Taurus}. 
These values correspond to opacities at $2.9~\mathrm{mm}$ of approximately $\kappa \approx 2.7, 1.5, 0.82, 0.45,$ and $0.25~\mathrm{cm^2~g^{-1}}$, respectively.

We calculated the flux density and spectral index using the following formula:
\begin{align}
F_\nu &= \int_{R_\mathrm{min}}^{R_\mathrm{max}}2\pi r \frac{B_\nu (1-e^{-\tau)}}{d^2}dr,\\
B_\nu &= \frac{2h\nu^3}{c^2}\frac{1}{\mathrm{exp}(\frac{h\nu}{kT})-1},\\
\alpha_\mathrm{mm} &= \frac{\ln(F_{341\mathrm{GHz}}/F_{105\mathrm{GHz}})}{\ln(341/105)},
\end{align}
\vspace{0.5em}
where $\tau = \kappa\Sigma$ is the optical depth and $d$ is the distance.

For each case, we calculate the flux at $140~\mathrm{pc}$ and the spectral index $\alpha_\mathrm{mm}$.
In all models, the disk is assumed to be face-on. 

\subsection{Result of model calculations}
\begin{figure*}
    \includegraphics[width=160mm]{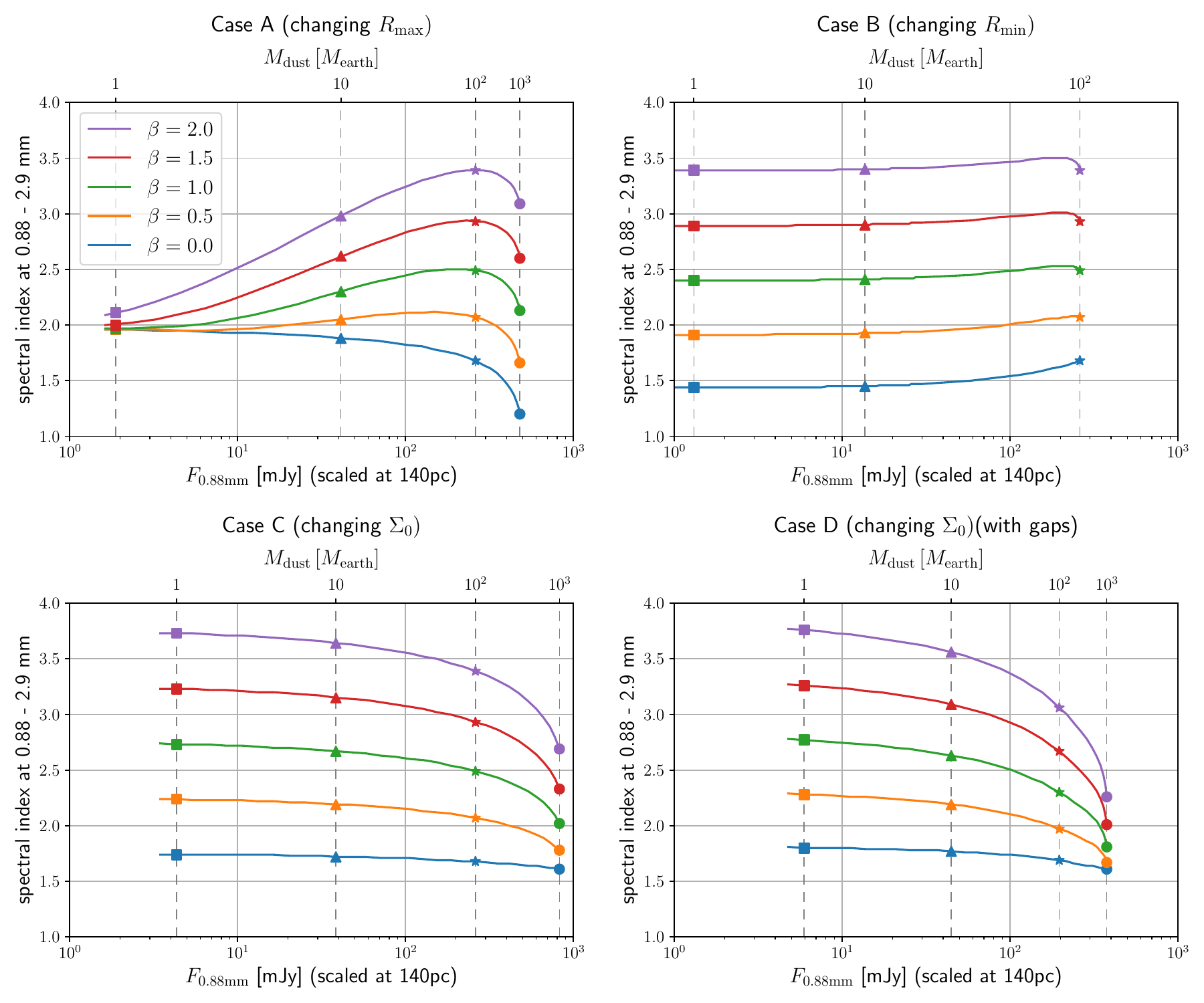}
    \caption{Calculated spectral indices between $0.88$ and $2.9~\mathrm{mm}$.
    The upper left panel shows case A, corresponding to a model with decreasing Rmax.
    The upper right panel shows case B, corresponding to a model with increasing Rmin.
    The lower panels show cases C and D, corresponding to models with decreasing $\Sigma_0$. case C (lower left) has no gap in the disk, while case D (lower right) has.
    In each panel, different colored lines indicate different values of $\beta$.
    The stars (\ensuremath{\star}) indicate the reference model, with $0.1~\mathrm{au}$ for the inner edge, $100~\mathrm{au}$ for the outer edge, and $100 M_\earth$ for the dust mass.
    Moving from one marker to the next corresponds to a tenfold change in dust mass.
    %{Alt text: Four panels of model curves for spectral index versus flux. Each panel shows five curves with $\beta = 0$, $0.5$, $1.0$, $1.5$, and $2.0$.}
    }
    \label{dust_model_x2}
\end{figure*}

Figure \ref{dust_model_x2} shows the result of the calculation.
In case A, the spectral index $\alpha_\mathrm{mm}$ rises once as $R_\mathrm{max}$ decreases and then approaches 2.
This is because in the larger disks, due to the lower temperatures in the outer part of the disk, the dust radiation was out of the Rayleigh-Jeans regime and $\alpha_\mathrm{mm}$ was smaller than $\beta+2$.
Then, the inner part of the disk is optically thick, thus approaching the spectral index of blackbody radiation.

In case B, since the inner part of the disk has a small area, its contribution to the total disk flux is small, and hence the total flux and $\alpha_\mathrm{mm}$ remain nearly constant as long as $R_\mathrm{min}\lesssim10$ au.
As $R_\mathrm{min}$ is increased beyond that, $\alpha_\mathrm{mm}$ approaches a value smaller than the expected $\beta + 2$ since only dust in optically thin and low-temperature region remains.
At the triangles (\ensuremath{\blacktriangle}), $R_\mathrm{min} = 90$ au, and optical depth is $\tau < 0.15$ in such outer region.

In cases C and D, $\alpha_\mathrm{mm}$ increases monotonically with decreasing surface density.
This is due to the extended optically thin region.
Compared to case C, case D has a little smaller flux when looking at the same dust mass. Also, the ratio of the change in $\alpha_\mathrm{mm}$ to the change in flux is larger.
However, there are no significant differences in the models with and without the gap.

\subsection{Evidence of grain growth}
\begin{figure*}
    \centering
    \includegraphics[width=160mm]{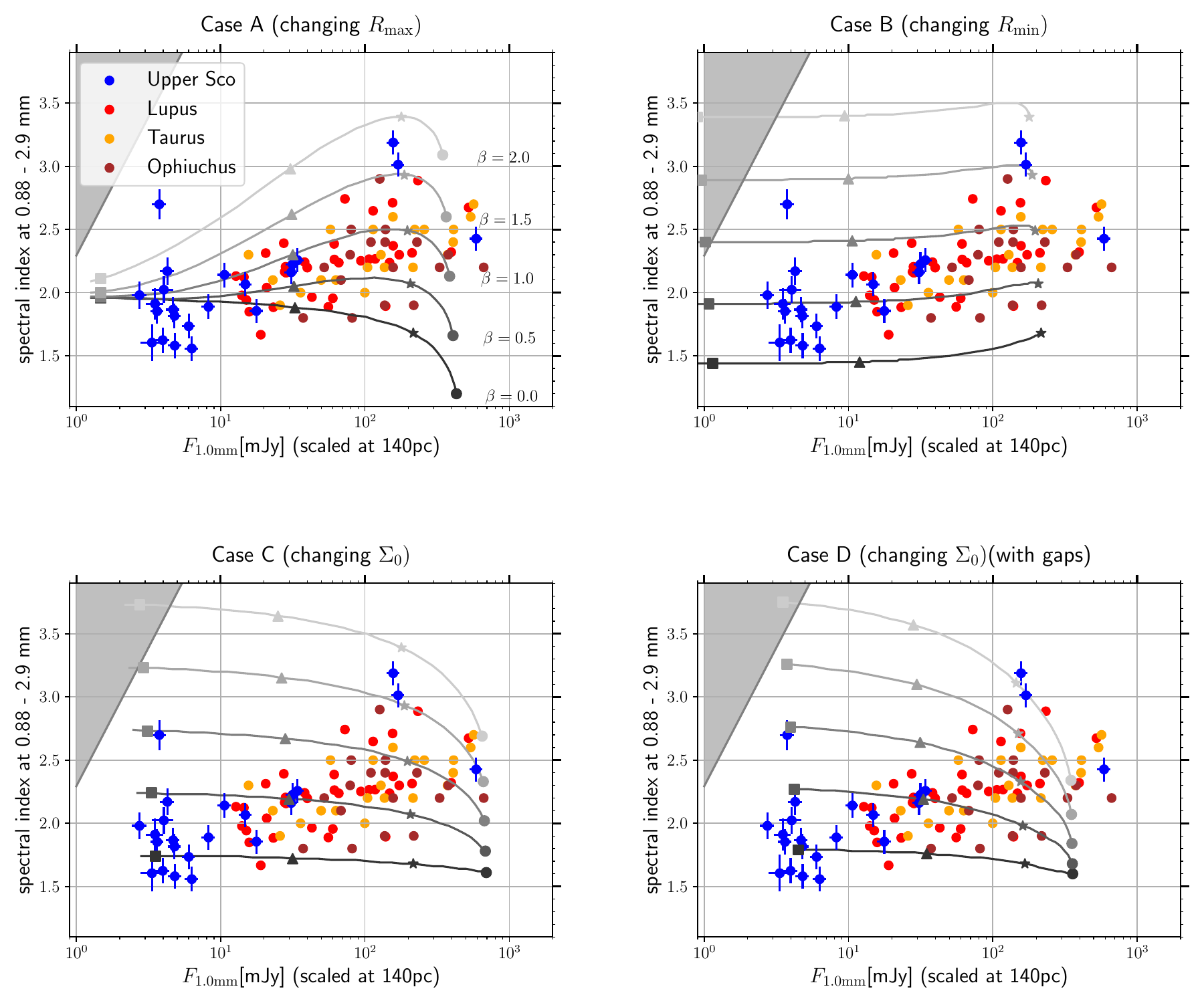}
    \caption{Observation vs. Model Comparison.
    Figures \ref{spectral_index} and \ref{dust_model_x2} are combined. For figure \ref{dust_model_x2}, the horizontal axis is converted from flux at $0.88~\mathrm{mm}$ to at $1 \mathrm{mm}$.
    %{Alt text: Four panels overlaying Figure \ref{dust_model_x2} model curves and Figure \ref{spectral_index} observational data.}
    }
    \label{model_obs_black}
\end{figure*}

Figure \ref{model_obs_black} compares the models with the observations.
Note that it is not our goal to explain the disk evolution with a single simple model. 
Rather, the aim is to investigate trends in how fluxes and spectral indices change depending on how mass loss proceeds, in order to constrain parameters such as disk radius.

The first important point is that no model can explain the USco observations, where $\alpha$ is less than 2, unless $\beta$ is sufficiently small.
Figure \ref{model_obs_black} suggests that, irrespective of the specific mass loss process, $\beta$ must be 0.5 or less in some of the faint USco disks.
As discussed in \citet{Ricci2010Taurus,Birnstiel2018,Tazzari2021}, $\beta$ is correlated with dust size.
Adopting the dust model from \citet{Ricci2010Taurus}, $\beta = 0.5$ corresponds to a maximum grain size of $3~\mathrm{mm}$ for $q = 2.5$, or $1~\mathrm{cm}$ for $q = 3.0$, where $q$ is the power law index of the grain size distribution, $n(a) \propto a^{-q}$.
While a degeneracy exists between the maximum grain size and the slope of size distribution, such low $\beta$ values can only be reproduced with both a flat dust size slope ($q\lesssim3$) and a large maximum grain size ($a_\mathrm{max}\gtrsim3~\mathrm{mm}$).
Upper Scorpius is much older (5-10 Myr; \citealp{Pecaut2016,Posch2024}) than the other three regions (1-3 Myr; \citealp{Alcala2017}).
Given this, at least for these USco objects, grain growth may have progressed further than in younger disks. 
For the other objects, a model with $\beta = 1$ can account for both young and old disks, making it difficult to determine whether dust grains grow further as disks evolve from younger to older stages.

Secondly, dust must be present in the cooler region of the disk.
If all the dust were in the high temperature region, $\alpha_\mathrm{mm}$ could not be less than 2 because Rayleigh-Jeans approximation would hold.
In case A, $\alpha$ approaches 2 because the dust will eventually be distributed only in the high-temperature region near the sun.
This model cannot explain the low values of $\alpha_\mathrm{mm}$ of USco disks.
To explain objects with $\alpha = 1.5$ and $\alpha = 1.75$ solely as a result of deviation from the Rayleigh–Jeans regime, the dust temperature must be 11.1 K and 20.9 K, respectively. 
These temperatures correspond to radial distances of $r = 81~\mathrm{au}$ and $r = 23~\mathrm{au}$ when the stellar luminosity is $0.25L_\odot$.
The models of uniformly decreasing surface density (case C and D) also fail to explain the extremely low $\alpha_\mathrm{mm}\sim 1.6$ even with the models with $\beta=0$.
Even when including the uncertainties, the observed data points lie below the $\beta=0$ model curves. This suggests that simple models in which the surface density uniformly decreases across the disk are not sufficient to explain the observations.
Only case B reproduces the small spectral indices, suggesting that dust must remain in the outer, cooler regions as the protoplanetary disk evolves.

We note that the actual disk evolution is expected to be much more complex than our model. A comparison between the data obtained in this study and more detailed disk population synthesis models (e.g., \citealt{Delussu2024}) is left for future work.

\subsection{Effect of non-thermal emission}
We showed that, under the assumption that the emission is purely thermal emission from dust, the presence of large dust in low-temperature regions is necessary to explain the low spectral indices observed in USco.
However, according to \citet{Hendler2020}, $R_{68}$ (radius containing 68\% of the total flux) of disks in USco is around 30 au. 
In such disks, it is unclear whether dust with temperature low enough to deviate from the Rayleigh-Jeans limit can dominate the disk emission.

Another possible explanation for the small spectral indices is non-thermal emission from ionized gas.
In the central regions near the star, free-free emission from ionized jets or disk winds is likely present \citep{CG16,Rota2024}. 
Here, we examine the amount of non-thermal emission required to explain the small spectral indices. 
Assuming that $F_{0.88\mathrm{mm}}$ is entirely due to thermal dust emission, we calculated the thermal emission component of $F_{2.9\mathrm{mm}}$ under the assumption of spectral indices of thermal dust emission $\alpha_\mathrm{thermal}=$ 2.0, 2.5, 3.0, and 3.5.
By comparing these results with the observation, we estimate the contribution of non-thermal emission to the total flux (Table \ref{nonthermal}).
The equations used for the calculations are as follows:
\begin{align}
F_{2.9\mathrm{mm,thermal}} &= F_{0.88\mathrm{mm,observed}}\left( \frac{0.88}{2.9}\right) ^{\alpha_\mathrm{thermal}},\\
F_{2.9\mathrm{mm,nonthermal}} &= F_{2.9\mathrm{mm,observed}} - F_{2.9\mathrm{mm,thermal}}.
\end{align}

Assuming $\alpha_\mathrm{thermal}=2.5$, the source J16135434-2320342, which has the lowest observed spectral index of 1.56, shows that the proportion of non-thermal emission is 67\% ($0.8~\mathrm{mJy}$ at $2.9~\mathrm{mm}$). 
For reference, the free-free emission of HL Tau, a much younger object with high accretion rate ($\sim10^{-7}M_{\odot}/{\rm yr}$; \citealp{White2004}), is estimated to be $\sim$ ${0.7~\mathrm{mJy}}$ at ${2.9~\mathrm{mm}}$ \citep{CG16}.
It is unclear whether older USco disks with lower accretion rates ($<10^{-8}M_{\odot}/{\rm yr}$; \citealp{Manara2020}) would exhibit a similar level of free-free emission.

Further observations at longer wavelengths are required to isolate the thermal emission from non-thermal contributions and to more accurately constrain its spectral index.
Recently, detailed microwave spectra covering $4\text{--}360~\mathrm{GHz}$ have been investigated for bright Taurus disks \citep{Painter2025}, enabling the quantification of free-free emission contamination at each frequency.
Extending such studies to USco disks is crucial for understanding the time evolution of protoplanetary disks.
In addition, high angular resolution data will allow us to probe regions of the disk where scattering and free-free emission are negligible, thereby providing better constraints on dust grain sizes, particularly in regions where thermal emission dominates.
Such observations may also uncover spatial variations in grain size across the disk.

\begin{longtable}{ccc|cccc}
\caption{Estimated proportion of non-thermal emission to total flux.}
\label{nonthermal}
\hline
Source & $F_{2.9\mathrm{mm}}$ &$\alpha_\mathrm{mm}$ & \multicolumn{4}{c}{Proportion of non-thermal emission to total flux at $2.9~\mathrm{mm}$}\\
&(mJy)&&$\alpha_\mathrm{thermal}=2.0$&$\alpha_\mathrm{thermal}=2.5$&$\alpha_\mathrm{thermal}=3.0$&$\alpha_\mathrm{thermal}=3.5$\\
\hline
\endhead
\hline
    \multicolumn{7}{p{0.8\textwidth}}{\textbf{Notes:}
    The proportion of non-thermal emission was calculated by assuming a spectral index for the thermal emission $\alpha_\mathrm{thermal}$ and comparing it with the observations at $2.9~\mathrm{mm}$.
    }\\
    \endlastfoot
J16113134-1838259&51.824&2.4&---&0.08&0.49&0.72\\
J16042165-2130284&5.122&3.2&---&---&---&0.31\\
J15583692-2257153&5.033&3.0&---&---&---&0.44\\
J16090075-1908526&3.313&2.3&---&0.25&0.58&0.77\\
J16082324-1930009&3.147&2.2&---&0.28&0.60&0.78\\
J16142029-1906481&3.193&2.2&---&0.33&0.63&0.79\\
J16154416-1921171&2.068&2.1&---&0.40&0.67&0.82\\
J16075796-2040087&2.643&1.9&0.16&0.53&0.74&0.86\\
J16072625-2432079&1.054&2.1&---&0.34&0.64&0.80\\
J16024152-2138245&1.109&1.9&0.12&0.51&0.73&0.85\\
J16054540-2023088&0.991&1.7&0.27&0.59&0.77&0.87\\
J16135434-2320342&1.202&1.6&0.41&0.67&0.82&0.90\\
J16123916-1859284&0.467&2.2&---&0.32&0.62&0.79\\
J15582981-2310077&0.651&1.9&0.15&0.53&0.74&0.85\\
J15530132-2114135&0.681&1.8&0.19&0.55&0.75&0.86\\
J16020757-2257467&0.219&2.7&---&---&0.30&0.61\\
J16111330-2019029&0.758&1.6&0.39&0.66&0.81&0.90\\
J16141107-2305362&0.705&1.6&0.36&0.64&0.80&0.89\\
J16181904-2028479&0.521&1.9&0.16&0.53&0.74&0.86\\
J16035767-2031055&0.453&1.9&0.10&0.50&0.72&0.85\\
J16062196-1928445&0.616&1.6&0.37&0.65&0.81&0.89\\
J16001844-2230114&0.359&2.0&---&0.43&0.68&0.82\\
J16014086-2258103&0.335&2.0&0.02&0.46&0.70&0.83\\
\end{longtable}

\section{Summary}
We analyzed ALMA Band 3 ($2.9~\mathrm{mm}$) observations of 23 disks in the Upper Scorpius star forming region and compared them with Band 7 ($0.88~\mathrm{mm}$) data \citep{Barenfeld2016} to calculate the spectral indices.
The observed spectral indices were compared with four simple parametric models of disk mass loss.
The key conclusions of this paper are as follows:
\begin{enumerate}
\item
The mean spectral index of the disks in USco is $\alpha_\mathrm{mm} = 2.1 \pm 0.1$, which is equal to or slightly smaller than those reported in younger regions such as Lupus ($\alpha_\mathrm{mm} \simeq 2.2$; \citealp{Tazzari2021}), Taurus ($\alpha_\mathrm{mm} \simeq 2.3$; \citealp{Ricci2010Taurus}), and Ophiuchus ($\alpha_\mathrm{mm} \simeq 2.2$; \citealp{Ricci2010Ophiuchus}).
\item
The observed small $\alpha_\mathrm{mm}$ values in USco disks indicate that grains have grown to a level where the millimeter opacity spectral index $\beta$ is below 0.5, even if other potential causes are considered in models such as optically thick regions and low-temperature regions.
\item 
The low values of $\alpha_\mathrm{mm} < 2$ can only be produced when the Rayleigh–Jeans approximation no longer applies. The tendency of such low values in these old disks suggests that the relative contribution of the cold outer disk must become more important as would be the case for inside-out clearing.
\end{enumerate}

\section*{Funding}
This work was supported by JSPS KAKENHI Grant Number JP22K03680.

T.B. acknowledges funding from the European Union under the European Union's Horizon Europe Research and Innovation Programme 101124282 (EARLYBIRD) and funding by the Deutsche Forschungsgemeinschaft (DFG, German Research Foundation) under grant 325594231, and Germany's Excellence Strategy - EXC-2094 - 390783311. Views and opinions expressed are, however, those of the authors only and do not necessarily reflect those of the European Union or the European Research Council. Neither the European Union nor the granting authority can be held responsible for them.

\section*{Acknowledgments}
This paper makes use of the following ALMA data: ADS/JAO.ALMA\#2015.1.00819.S. ALMA is a partnership of ESO (representing its member states), NSF (USA) and NINS (Japan), together with NRC (Canada), NSTC and ASIAA (Taiwan), and KASI (Republic of Korea), in cooperation with the Republic of Chile. The Joint ALMA Observatory is operated by ESO, AUI/NRAO and NAOJ.

\section*{Appendix. Distances and corrected fluxes of disks in Lupus, Taurus, and Ophiuchus}
\label{distances}
To calculate absolute disk fluxes, we used the parallax data \citep{GaiaDR3} and calculate the distances of disks.
Tables \ref{appendix_Taurus}, \ref{appendix_Ophiuchus}, and \ref{appendix_Lupus} show the parallaxes, distances, and fluxes at $1~\mathrm{mm}$ scaled at $140~\mathrm{pc}$.

\begin{table}
\caption{Distances and corrected fluxes of disks in Taurus.}
\label{appendix_Taurus}
\begin{tabular}{*{4}{c}}
\hline
Source & parallax & distance & $F_{1\mathrm{mm}, 140~\mathrm{pc}}$\\
&(marcsec)&(pc)&(mJy)\\
\hline
AA Tau&$7.43\pm{0.09}$&$134.6^{+1.7}_{-1.6}$&$99.8^{+2.5}_{-2.4}$\\
CI Tau&$6.24\pm{0.02}$&$160.3^{+0.5}_{-0.5}$&$411.4^{+2.7}_{-2.6}$\\
CW Tau&$7.60\pm{0.04}$&$131.6^{+0.7}_{-0.7}$&$113.9^{+1.2}_{-1.2}$\\
CX Tau&$7.89\pm{0.02}$&$126.7^{+0.3}_{-0.3}$&$15.6^{+0.1}_{-0.1}$\\
CY Tau&$7.92\pm{0.02}$&$126.3^{+0.3}_{-0.3}$&$136.6^{+0.7}_{-0.7}$\\
DE Tau&$7.81\pm{0.03}$&$128.0^{+0.5}_{-0.5}$&$57.7^{+0.4}_{-0.4}$\\
DL Tau&$6.25\pm{0.02}$&$160.0^{+0.5}_{-0.5}$&$408.8^{+2.6}_{-2.6}$\\
DM Tau&$6.94\pm{0.02}$&$144.1^{+0.4}_{-0.4}$&$221.4^{+1.3}_{-1.3}$\\
DN Tau&$7.78\pm{0.02}$&$128.5^{+0.3}_{-0.3}$&$129.0^{+0.7}_{-0.7}$\\
DO Tau&$7.22\pm{0.04}$&$138.5^{+0.8}_{-0.8}$&$215.3^{+2.4}_{-2.4}$\\
DR Tau&$5.18\pm{0.03}$&$193.1^{+1.1}_{-1.1}$&$566.6^{+6.6}_{-6.5}$\\
DS Tau&$6.32\pm{0.02}$&$158.2^{+0.5}_{-0.5}$&$35.8^{+0.2}_{-0.2}$\\
FM Tau&$7.58\pm{0.02}$&$131.9^{+0.3}_{-0.3}$&$25.8^{+0.1}_{-0.1}$\\
FZ Tau&$7.74\pm{0.03}$&$129.2^{+0.5}_{-0.5}$&$23.0^{+0.2}_{-0.2}$\\
GM Aur&$6.32\pm{0.05}$&$158.2^{+1.3}_{-1.2}$&$540.3^{+8.7}_{-8.4}$\\
GO Tau&$7.02\pm{0.02}$&$142.5^{+0.4}_{-0.4}$&$156.3^{+0.9}_{-0.9}$\\
HO Tau&$6.08\pm{0.03}$&$164.5^{+0.8}_{-0.8}$&$49.7^{+0.5}_{-0.5}$\\
IQ Tau&$7.60\pm{0.04}$&$131.6^{+0.7}_{-0.7}$&$104.2^{+1.1}_{-1.1}$\\
RY Tau&$7.23\pm{0.20}$&$138.3^{+3.9}_{-3.7}$&$373.8^{+21.6}_{-19.9}$\\
SU Aur&$6.37\pm{0.03}$&$157.0^{+0.7}_{-0.7}$&$62.9^{+0.6}_{-0.6}$\\
UZ Tau E&$8.12\pm{0.45}$&$123.2^{+7.2}_{-6.5}$&$257.7^{+31.1}_{-26.4}$\\
\hline
\end{tabular}
\parbox{0.45\textwidth}{\small
\textbf{Note:} The parallaxes are from Gaia DR3 measurements \citep{GaiaDR3}, and the distances are computed using them. 
The scaled fluxes are calculated using data from \citet{Ricci2010Taurus}.
}
\end{table}

\begin{table}
\caption{Distances and corrected fluxes of disks in Ophiuchus.}
\label{appendix_Ophiuchus}
\begin{tabular}{*{4}{c}}
\hline
Source & parallax & distance & $F_{1\mathrm{mm}, 140 \mathrm{pc}}$\\
&(marcsec)&(pc)&(mJy)\\
\hline
SR 4&$7.42\pm{0.03}$&$134.8\pm{0.5}$&$80.3^{+0.7}_{-0.6}$\\
GSS 26&$7.39\pm{0.02}$&$135.3\pm{0.4}$&$217.8\pm{1.2}$\\
EL 20&$7.27\pm{0.21}$&$137.6^{+4.1}_{-3.9}$&$138.8^{+8.4}_{-7.7}$\\
DoAr 25&$7.24\pm{0.04}$&$138.1\pm{0.8}$&$377.7^{+4.2}_{-4.1}$\\
EL 24&$7.18\pm{0.06}$&$139.3\pm{1.2}$&$665.3^{+11.3}_{-11.0}$\\
EL 27&$9.09\pm{0.85}$&$110.0^{+11.3}_{-9.4}$&$329.2^{+71.4}_{-53.9}$\\
SR 21&$7.33\pm{0.03}$&$136.4\pm{0.6}$&$126.6\pm{1.0}$\\
YLW 16c&$7.26\pm{0.08}$&$137.7\pm{1.5}$&$110.0^{+2.5}_{-2.4}$\\
IRS 49&$7.17\pm{0.13}$&$139.5^{+2.6}_{-2.5}$&$37.3^{+1.4}_{-1.3}$\\
DoAr 33&$7.06\pm{0.02}$&$141.6\pm{0.4}$&$52.2\pm{0.3}$\\
WSB 52&$7.39\pm{0.05}$&$135.3\pm{0.9}$&$81.5\pm{1.1}$\\
WSB 60&$7.41\pm{0.08}$&$135.0^{+1.5}_{-1.4}$&$137.0^{+3.0}_{-2.9}$\\
DoAr 44&$6.83\pm{0.02}$&$146.4\pm{0.4}$&$156.7\pm{0.9}$\\
RNO 90&$8.73\pm{0.03}$&$114.5\pm{0.4}$&$79.0\pm{0.5}$\\
Wa Oph 6&$8.16\pm{0.02}$&$122.5\pm{0.3}$&$138.0\pm{0.7}$\\
AS 209&$8.25\pm{0.03}$&$121.2\pm{0.4}$&$229.4\pm{1.7}$\\
\hline
\end{tabular}
\parbox{0.45\textwidth}{\small
\textbf{Note:} The parallaxes are from Gaia DR3 measurements \citep{GaiaDR3}, and the distances are computed using them. 
The scaled fluxes are calculated using data from \citet{Ricci2010Ophiuchus}.
}
\end{table}

\begin{table}
\caption{Distances and corrected fluxes of disks in Lupus.}
\label{appendix_Lupus}
\begin{tabular}{*{4}{c}}
\hline
Source & parallax & distance & $F_{1\mathrm{mm}, 140 \mathrm{pc}}$\\
&(marcsec)&(pc)&(mJy)\\
\hline
Sz 65&$6.52\pm{0.03}$&$153.4\pm{0.7}$&$58.9\pm{0.5}$\\
Sz 66&$6.41\pm{0.03}$&$156.0\pm{0.7}$&$13.8\pm{0.1}$\\
J15450634-3417378&$6.46\pm{0.14}$&$154.8^{+3.4}_{-3.3}$&$13.6\pm{0.6}$\\
J15450887-3417333&$6.46\pm{0.14}$&$154.8^{+3.4}_{-3.3}$&$42.5^{+1.9}_{-1.8}$\\
Sz 68&$6.52\pm{0.05}$&$153.4\pm{1.2}$&$137.4\pm{2.1}$\\
Sz 69&$6.55\pm{0.07}$&$152.7\pm{1.6}$&$15.5\pm{0.3}$\\
Sz 71&$6.44\pm{0.02}$&$155.3\pm{0.5}$&$149.0\pm{0.9}$\\
Sz 72&$6.38\pm{0.02}$&$156.7\pm{0.5}$&$13.0\pm{0.1}$\\
Sz 73&$6.34\pm{0.03}$&$157.7\pm{0.7}$&$28.2\pm{0.3}$\\
Sz 74&$7.06\pm{0.39}$&$141.6^{+8.3}_{-7.4}$&$17.0^{+2.0}_{-1.7}$\\
Sz 75&$6.49\pm{0.03}$&$154.1\pm{0.7}$&$68.7\pm{0.6}$\\
Sz 82&$6.42\pm{0.02}$&$155.8\pm{0.5}$&$505.9\pm{3.2}$\\
Sz 83&$6.35\pm{0.04}$&$157.5\pm{1.0}$&$388.3\pm{4.9}$\\
Sz 84&$6.34\pm{0.04}$&$157.7\pm{1.0}$&$29.5\pm{0.4}$\\
Sz 129&$6.24\pm{0.02}$&$160.3\pm{0.5}$&$171.2\pm{1.1}$\\
J15592838-4021513&$6.52\pm{0.06}$&$153.4\pm{1.4}$&$220.4\pm{4.1}$\\
J16000236-4222145&$6.23\pm{0.04}$&$160.5\pm{1.0}$&$114.5\pm{1.5}$\\
J16004452-4155310&$6.36\pm{0.04}$&$157.2\pm{1.0}$&$159.0\pm{2.0}$\\
Sz 133 &$8.62\pm{0.62}$&$116.0^{+9.0}_{-7.8}$&$34.3^{+5.5}_{-4.4}$\\
J16070854-3914075&$5.84\pm{0.34}$&$171.2^{+10.6}_{-9.4}$&$99.3^{+12.7}_{-10.6}$\\
Sz 90&$6.24\pm{0.02}$&$160.3\pm{0.5}$&$20.5\pm{0.1}$\\
Sz 91&$6.27\pm{0.03}$&$159.5\pm{0.8}$&---\\
Sz 98&$6.40\pm{0.02}$&$156.3\pm{0.5}$&$211.6\pm{1.3}$\\
Sz 100&$7.09\pm{0.13}$&$141.0^{+2.6}_{-2.5}$&$40.4^{+1.5}_{-1.4}$\\
J16083070-3828268 &$6.52\pm{0.03}$&$153.4\pm{0.7}$&$111.9\pm{1.0}$\\
J16083427-3906181&$6.31\pm{0.03}$&$158.5^{+0.8}_{-0.7}$&$54.9\pm{0.5}$\\
Sz 108B&$6.20\pm{0.06}$&$161.3^{+1.6}_{-1.5}$&$26.2\pm{0.5}$\\
Sz 110&$6.35\pm{0.02}$&$157.5\pm{0.5}$&$14.7\pm{0.1}$\\
J16085324-3914401 &$6.13\pm{0.05}$&$163.1\pm{1.3}$&$19.9\pm{0.3}$\\
Sz 111&$6.31\pm{0.02}$&$158.5\pm{0.5}$&$156.1\pm{1.0}$\\
Sz 113&$6.23\pm{0.05}$&$160.5\pm{1.3}$&$22.4\pm{0.4}$\\
Sz 114&$6.38\pm{0.02}$&$156.7\pm{0.5}$&$87.6^{+0.6}_{-0.5}$\\
Sz 118&$6.19\pm{0.03}$&$161.6\pm{0.8}$&$60.4\pm{0.6}$\\
Sz 123&$6.17\pm{0.02}$&$162.1\pm{0.5}$&$40.0\pm{0.3}$\\
J16124373-3815031&$6.26\pm{0.02}$&$159.7\pm{0.5}$&$28.3\pm{0.2}$\\
\hline
\end{tabular}
\parbox{0.45\textwidth}{\small
\textbf{Note:} The parallaxes are from Gaia DR3 measurements \citep{GaiaDR3}, and the distances are computed using them. 
The scaled fluxes are calculated using data from \citet{Tazzari2021}.
Sz 91 was not detected due to an erroneous observational setup.
}
\end{table}

\bibliography{papers}

\begin{thebibliography}{}
\expandafter\ifx\csname natexlab\endcsname\relax\def\natexlab#1{#1}\fi

\bibitem[{{Alcal{\'a}} {et~al.}(2017){Alcal{\'a}}, {Manara}, {Natta}, {Frasca}, {Testi}, {Nisini}, {Stelzer}, {Williams}, {Antoniucci}, {Biazzo}, {Covino}, {Esposito}, {Getman}, \& {Rigliaco}}]{Alcala2017}
{Alcal{\'a}}, J.~M., {Manara}, C.~F., {Natta}, A., {et~al.} 2017, \aap, 600, A20

\bibitem[{{Andre} {et~al.}(1993){Andre}, {Ward-Thompson}, \& {Barsony}}]{Andre1993}
{Andre}, P., {Ward-Thompson}, D., \& {Barsony}, M. 1993, \apj, 406, 122

\bibitem[{{Andrews}(2020)}]{Andrews2020}
{Andrews}, S.~M. 2020, \araa, 58, 483

\bibitem[{{Andrews} {et~al.}(2018){Andrews}, {Huang}, {P{\'e}rez}, {Isella}, {Dullemond}, {Kurtovic}, {Guzm{\'a}n}, {Carpenter}, {Wilner}, {Zhang}, {Zhu}, {Birnstiel}, {Bai}, {Benisty}, {Hughes}, {{\"O}berg}, \& {Ricci}}]{Andrews2018}
{Andrews}, S.~M., {Huang}, J., {P{\'e}rez}, L.~M., {et~al.} 2018, \apjl, 869, L41

\bibitem[{{Ansdell} {et~al.}(2016){Ansdell}, {Williams}, {van der Marel}, {Carpenter}, {Guidi}, {Hogerheijde}, {Mathews}, {Manara}, {Miotello}, {Natta}, {Oliveira}, {Tazzari}, {Testi}, {van Dishoeck}, \& {van Terwisga}}]{Ansdell2016}
{Ansdell}, M., {Williams}, J.~P., {van der Marel}, N., {et~al.} 2016, \apj, 828, 46

\bibitem[{{Barenfeld} {et~al.}(2016){Barenfeld}, {Carpenter}, {Ricci}, \& {Isella}}]{Barenfeld2016}
{Barenfeld}, S.~A., {Carpenter}, J.~M., {Ricci}, L., \& {Isella}, A. 2016, \apj, 827, 142

\bibitem[{{Beckwith} \& {Sargent}(1991)}]{Beckwith1991}
{Beckwith}, S. V.~W., \& {Sargent}, A.~I. 1991, \apj, 381, 250

\bibitem[{{Birnstiel}(2024)}]{Birnstiel2024}
{Birnstiel}, T. 2024, \araa, 62, 157

\bibitem[{{Birnstiel} {et~al.}(2018){Birnstiel}, {Dullemond}, {Zhu}, {Andrews}, {Bai}, {Wilner}, {Carpenter}, {Huang}, {Isella}, {Benisty}, {P{\'e}rez}, \& {Zhang}}]{Birnstiel2018}
{Birnstiel}, T., {Dullemond}, C.~P., {Zhu}, Z., {et~al.} 2018, \apjl, 869, L45

\bibitem[{{Carrasco-Gonz{\'a}lez} {et~al.}(2016){Carrasco-Gonz{\'a}lez}, {Henning}, {Chandler}, {Linz}, {P{\'e}rez}, {Rodr{\'\i}guez}, {Galv{\'a}n-Madrid}, {Anglada}, {Birnstiel}, {van Boekel}, {Flock}, {Klahr}, {Macias}, {Menten}, {Osorio}, {Testi}, {Torrelles}, \& {Zhu}}]{CG16}
{Carrasco-Gonz{\'a}lez}, C., {Henning}, T., {Chandler}, C.~J., {et~al.} 2016, \apjl, 821, L16

\bibitem[{{Carrasco-Gonz{\'a}lez} {et~al.}(2019){Carrasco-Gonz{\'a}lez}, {Sierra}, {Flock}, {Zhu}, {Henning}, {Chandler}, {Galv{\'a}n-Madrid}, {Mac{\'\i}as}, {Anglada}, {Linz}, {Osorio}, {Rodr{\'\i}guez}, {Testi}, {Torrelles}, {P{\'e}rez}, \& {Liu}}]{CG19}
{Carrasco-Gonz{\'a}lez}, C., {Sierra}, A., {Flock}, M., {et~al.} 2019, \apj, 883, 71

\bibitem[{{Delussu} {et~al.}(2024){Delussu}, {Birnstiel}, {Miotello}, {Pinilla}, {Rosotti}, \& {Andrews}}]{Delussu2024}
{Delussu}, L., {Birnstiel}, T., {Miotello}, A., {et~al.} 2024, \aap, 688, A81

\bibitem[{{Dipierro} {et~al.}(2015){Dipierro}, {Price}, {Laibe}, {Hirsh}, {Cerioli}, \& {Lodato}}]{Dipierro2015}
{Dipierro}, G., {Price}, D., {Laibe}, G., {et~al.} 2015, \mnras, 453, L73

\bibitem[{{Doi} \& {Kataoka}(2023)}]{Doi2023}
{Doi}, K., \& {Kataoka}, A. 2023, \apj, 957, 11

\bibitem[{{Draine}(2006)}]{Draine2006}
{Draine}, B.~T. 2006, \apj, 636, 1114

\bibitem[{{Espaillat} {et~al.}(2012){Espaillat}, {Ingleby}, {Hern{\'a}ndez}, {Furlan}, {D'Alessio}, {Calvet}, {Andrews}, {Muzerolle}, {Qi}, \& {Wilner}}]{Espaillat2012}
{Espaillat}, C., {Ingleby}, L., {Hern{\'a}ndez}, J., {et~al.} 2012, \apj, 747, 103

\bibitem[{{Finkbeiner} {et~al.}(1999){Finkbeiner}, {Davis}, \& {Schlegel}}]{Finkbeiner1999}
{Finkbeiner}, D.~P., {Davis}, M., \& {Schlegel}, D.~J. 1999, \apj, 524, 867

\bibitem[{{Flock} {et~al.}(2015){Flock}, {Ruge}, {Dzyurkevich}, {Henning}, {Klahr}, \& {Wolf}}]{Flock2015}
{Flock}, M., {Ruge}, J.~P., {Dzyurkevich}, N., {et~al.} 2015, \aap, 574, A68

\bibitem[{{Gaia Collaboration} {et~al.}(2023){Gaia Collaboration}, {Vallenari}, {Brown}, {Prusti}, {de Bruijne}, {Arenou}, {Babusiaux}, {Biermann}, {Creevey}, {Ducourant}, {Evans}, {Eyer}, {Guerra}, {Hutton}, {Jordi}, {Klioner}, {Lammers}, {Lindegren}, {Luri}, {Mignard}, {Panem}, {Pourbaix}, {Randich}, {Sartoretti}, {Soubiran}, {Tanga}, {Walton}, {Bailer-Jones}, {Bastian}, {Drimmel}, {Jansen}, {Katz}, {Lattanzi}, {van Leeuwen}, {Bakker}, {Cacciari}, {Casta{\~n}eda}, {De Angeli}, {Fabricius}, {Fouesneau}, {Fr{\'e}mat}, {Galluccio}, {Guerrier}, {Heiter}, {Masana}, {Messineo}, {Mowlavi}, {Nicolas}, {Nienartowicz}, {Pailler}, {Panuzzo}, {Riclet}, {Roux}, {Seabroke}, {Sordo}, {Th{\'e}venin}, {Gracia-Abril}, {Portell}, {Teyssier}, {Altmann}, {Andrae}, {Audard}, {Bellas-Velidis}, {Benson}, {Berthier}, {Blomme}, {Burgess}, {Busonero}, {Busso}, {C{\'a}novas}, {Carry}, {Cellino}, {Cheek}, {Clementini}, {Damerdji}, {Davidson}, {de Teodoro}, {Nu{\~n}ez Campos}, {Delchambre}, {Dell'Oro}, {Esquej},
  {Fern{\'a}ndez-Hern{\'a}ndez}, {Fraile}, {Garabato}, {Garc{\'\i}a-Lario}, {Gosset}, {Haigron}, {Halbwachs}, {Hambly}, {Harrison}, {Hern{\'a}ndez}, {Hestroffer}, {Hodgkin}, {Holl}, {Jan{\ss}en}, {Jevardat de Fombelle}, {Jordan}, {Krone-Martins}, {Lanzafame}, {L{\"o}ffler}, {Marchal}, {Marrese}, {Moitinho}, {Muinonen}, {Osborne}, {Pancino}, {Pauwels}, {Recio-Blanco}, {Reyl{\'e}}, {Riello}, {Rimoldini}, {Roegiers}, {Rybizki}, {Sarro}, {Siopis}, {Smith}, {Sozzetti}, {Utrilla}, {van Leeuwen}, {Abbas}, {{\'A}brah{\'a}m}, {Abreu Aramburu}, {Aerts}, {Aguado}, {Ajaj}, {Aldea-Montero}, {Altavilla}, {{\'A}lvarez}, {Alves}, {Anders}, {Anderson}, {Anglada Varela}, {Antoja}, {Baines}, {Baker}, {Balaguer-N{\'u}{\~n}ez}, {Balbinot}, {Balog}, {Barache}, {Barbato}, {Barros}, {Barstow}, {Bartolom{\'e}}, {Bassilana}, {Bauchet}, {Becciani}, {Bellazzini}, {Berihuete}, {Bernet}, {Bertone}, {Bianchi}, {Binnenfeld}, {Blanco-Cuaresma}, {Blazere}, {Boch}, {Bombrun}, {Bossini}, {Bouquillon}, {Bragaglia}, {Bramante}, {Breedt},
  {Bressan}, {Brouillet}, {Brugaletta}, {Bucciarelli}, {Burlacu}, {Butkevich}, {Buzzi}, {Caffau}, {Cancelliere}, {Cantat-Gaudin}, {Carballo}, {Carlucci}, {Carnerero}, {Carrasco}, {Casamiquela}, {Castellani}, {Castro-Ginard}, {Chaoul}, {Charlot}, {Chemin}, {Chiaramida}, {Chiavassa}, {Chornay}, {Comoretto}, {Contursi}, {Cooper}, {Cornez}, {Cowell}, {Crifo}, {Cropper}, {Crosta}, {Crowley}, {Dafonte}, {Dapergolas}, {David}, {David}, {de Laverny}, {De Luise}, {De March}, {De Ridder}, {de Souza}, {de Torres}, {del Peloso}, {del Pozo}, {Delbo}, {Delgado}, {Delisle}, {Demouchy}, {Dharmawardena}, {Di Matteo}, {Diakite}, {Diener}, {Distefano}, {Dolding}, {Edvardsson}, {Enke}, {Fabre}, {Fabrizio}, {Faigler}, {Fedorets}, {Fernique}, {Fienga}, {Figueras}, {Fournier}, {Fouron}, {Fragkoudi}, {Gai}, {Garcia-Gutierrez}, {Garcia-Reinaldos}, {Garc{\'\i}a-Torres}, {Garofalo}, {Gavel}, {Gavras}, {Gerlach}, {Geyer}, {Giacobbe}, {Gilmore}, {Girona}, {Giuffrida}, {Gomel}, {Gomez}, {Gonz{\'a}lez-N{\'u}{\~n}ez},
  {Gonz{\'a}lez-Santamar{\'\i}a}, {Gonz{\'a}lez-Vidal}, {Granvik}, {Guillout}, {Guiraud}, {Guti{\'e}rrez-S{\'a}nchez}, {Guy}, {Hatzidimitriou}, {Hauser}, {Haywood}, {Helmer}, {Helmi}, {Sarmiento}, {Hidalgo}, {Hilger}, {H{\l}adczuk}, {Hobbs}, {Holland}, {Huckle}, {Jardine}, {Jasniewicz}, {Jean-Antoine Piccolo}, {Jim{\'e}nez-Arranz}, {Jorissen}, {Juaristi Campillo}, {Julbe}, {Karbevska}, {Kervella}, {Khanna}, {Kontizas}, {Kordopatis}, {Korn}, {K{\'o}sp{\'a}l}, {Kostrzewa-Rutkowska}, {Kruszy{\'n}ska}, {Kun}, {Laizeau}, {Lambert}, {Lanza}, {Lasne}, {Le Campion}, {Lebreton}, {Lebzelter}, {Leccia}, {Leclerc}, {Lecoeur-Taibi}, {Liao}, {Licata}, {Lindstr{\o}m}, {Lister}, {Livanou}, {Lobel}, {Lorca}, {Loup}, {Madrero Pardo}, {Magdaleno Romeo}, {Managau}, {Mann}, {Manteiga}, {Marchant}, {Marconi}, {Marcos}, {Marcos Santos}, {Mar{\'\i}n Pina}, {Marinoni}, {Marocco}, {Marshall}, {Martin Polo}, {Mart{\'\i}n-Fleitas}, {Marton}, {Mary}, {Masip}, {Massari}, {Mastrobuono-Battisti}, {Mazeh}, {McMillan}, {Messina}, {Michalik},
  {Millar}, {Mints}, {Molina}, {Molinaro}, {Moln{\'a}r}, {Monari}, {Mongui{\'o}}, {Montegriffo}, {Montero}, {Mor}, {Mora}, {Morbidelli}, {Morel}, {Morris}, {Muraveva}, {Murphy}, {Musella}, {Nagy}, {Noval}, {Oca{\~n}a}, {Ogden}, {Ordenovic}, {Osinde}, {Pagani}, {Pagano}, {Palaversa}, {Palicio}, {Pallas-Quintela}, {Panahi}, {Payne-Wardenaar}, {Pe{\~n}alosa Esteller}, {Penttil{\"a}}, {Pichon}, {Piersimoni}, {Pineau}, {Plachy}, {Plum}, {Poggio}, {Pr{\v{s}}a}, {Pulone}, {Racero}, {Ragaini}, {Rainer}, {Raiteri}, {Rambaux}, {Ramos}, {Ramos-Lerate}, {Re Fiorentin}, {Regibo}, {Richards}, {Rios Diaz}, {Ripepi}, {Riva}, {Rix}, {Rixon}, {Robichon}, {Robin}, {Robin}, {Roelens}, {Rogues}, {Rohrbasser}, {Romero-G{\'o}mez}, {Rowell}, {Royer}, {Ruz Mieres}, {Rybicki}, {Sadowski}, {S{\'a}ez N{\'u}{\~n}ez}, {Sagrist{\`a} Sell{\'e}s}, {Sahlmann}, {Salguero}, {Samaras}, {Sanchez Gimenez}, {Sanna}, {Santove{\~n}a}, {Sarasso}, {Schultheis}, {Sciacca}, {Segol}, {Segovia}, {S{\'e}gransan}, {Semeux}, {Shahaf}, {Siddiqui}, {Siebert},
  {Siltala}, {Silvelo}, {Slezak}, {Slezak}, {Smart}, {Snaith}, {Solano}, {Solitro}, {Souami}, {Souchay}, {Spagna}, {Spina}, {Spoto}, {Steele}, {Steidelm{\"u}ller}, {Stephenson}, {S{\"u}veges}, {Surdej}, {Szabados}, {Szegedi-Elek}, {Taris}, {Taylor}, {Teixeira}, {Tolomei}, {Tonello}, {Torra}, {Torra}, {Torralba Elipe}, {Trabucchi}, {Tsounis}, {Turon}, {Ulla}, {Unger}, {Vaillant}, {van Dillen}, {van Reeven}, {Vanel}, {Vecchiato}, {Viala}, {Vicente}, {Voutsinas}, {Weiler}, {Wevers}, {Wyrzykowski}, {Yoldas}, {Yvard}, {Zhao}, {Zorec}, {Zucker}, \& {Zwitter}}]{GaiaDR3}
{Gaia Collaboration}, {Vallenari}, A., {Brown}, A.~G.~A., {et~al.} 2023, \aap, 674, A1

\bibitem[{{Guidi} {et~al.}(2022){Guidi}, {Isella}, {Testi}, {Chandler}, {Liu}, {Schmid}, {Rosotti}, {Meng}, {Jennings}, {Williams}, {Carpenter}, {de Gregorio-Monsalvo}, {Li}, {Liu}, {Ortolani}, {Quanz}, {Ricci}, \& {Tazzari}}]{Guidi2022}
{Guidi}, G., {Isella}, A., {Testi}, L., {et~al.} 2022, \aap, 664, A137

\bibitem[{{Hayashi}(1981)}]{Hayashi1981}
{Hayashi}, C. 1981, Progress of Theoretical Physics Supplement, 70, 35

\bibitem[{{Hendler} {et~al.}(2020){Hendler}, {Pascucci}, {Pinilla}, {Tazzari}, {Carpenter}, {Malhotra}, \& {Testi}}]{Hendler2020}
{Hendler}, N., {Pascucci}, I., {Pinilla}, P., {et~al.} 2020, \apj, 895, 126

\bibitem[{{Huang} {et~al.}(2018){Huang}, {Andrews}, {Dullemond}, {Isella}, {P{\'e}rez}, {Guzm{\'a}n}, {{\"O}berg}, {Zhu}, {Zhang}, {Bai}, {Benisty}, {Birnstiel}, {Carpenter}, {Hughes}, {Ricci}, {Weaver}, \& {Wilner}}]{Huang2018}
{Huang}, J., {Andrews}, S.~M., {Dullemond}, C.~P., {et~al.} 2018, \apjl, 869, L42

\bibitem[{{Lada} \& {Wilking}(1984)}]{Lada1984}
{Lada}, C.~J., \& {Wilking}, B.~A. 1984, \apj, 287, 610

\bibitem[{{Law} {et~al.}(2021){Law}, {Teague}, {Loomis}, {Bae}, {{\"O}berg}, {Czekala}, {Andrews}, {Aikawa}, {Alarc{\'o}n}, {Bergin}, {Bergner}, {Booth}, {Bosman}, {Calahan}, {Cataldi}, {Cleeves}, {Furuya}, {Guzm{\'a}n}, {Huang}, {Ilee}, {Le Gal}, {Liu}, {Long}, {M{\'e}nard}, {Nomura}, {P{\'e}rez}, {Qi}, {Schwarz}, {Soto}, {Tsukagoshi}, {Yamato}, {van't Hoff}, {Walsh}, {Wilner}, \& {Zhang}}]{Law2021}
{Law}, C.~J., {Teague}, R., {Loomis}, R.~A., {et~al.} 2021, \apjs, 257, 4

\bibitem[{{Luhman} \& {Esplin}(2020)}]{Luhman2020}
{Luhman}, K.~L., \& {Esplin}, T.~L. 2020, \aj, 160, 44

\bibitem[{{Luhman} \& {Mamajek}(2012)}]{LuhmanMamajek2012}
{Luhman}, K.~L., \& {Mamajek}, E.~E. 2012, \apj, 758, 31

\bibitem[{{Lynden-Bell} \& {Pringle}(1974)}]{Lyden1974}
{Lynden-Bell}, D., \& {Pringle}, J.~E. 1974, \mnras, 168, 603

\bibitem[{{Mac{\'\i}as} {et~al.}(2021){Mac{\'\i}as}, {Guerra-Alvarado}, {Carrasco-Gonz{\'a}lez}, {Ribas}, {Espaillat}, {Huang}, \& {Andrews}}]{Macias2021}
{Mac{\'\i}as}, E., {Guerra-Alvarado}, O., {Carrasco-Gonz{\'a}lez}, C., {et~al.} 2021, \aap, 648, A33

\bibitem[{{Manara} {et~al.}(2020){Manara}, {Natta}, {Rosotti}, {Alcal{\'a}}, {Nisini}, {Lodato}, {Testi}, {Pascucci}, {Hillenbrand}, {Carpenter}, {Scholz}, {Fedele}, {Frasca}, {Mulders}, {Rigliaco}, {Scardoni}, \& {Zari}}]{Manara2020}
{Manara}, C.~F., {Natta}, A., {Rosotti}, G.~P., {et~al.} 2020, \aap, 639, A58

\bibitem[{{McMullin} {et~al.}(2007){McMullin}, {Waters}, {Schiebel}, {Young}, \& {Golap}}]{McMullin2007}
{McMullin}, J.~P., {Waters}, B., {Schiebel}, D., {Young}, W., \& {Golap}, K. 2007, in Astronomical Society of the Pacific Conference Series, Vol. 376, Astronomical Data Analysis Software and Systems XVI, ed. R.~A. {Shaw}, F.~{Hill}, \& D.~J. {Bell}, 127

\bibitem[{{Miyake} \& {Nakagawa}(1993)}]{Miyake1993}
{Miyake}, K., \& {Nakagawa}, Y. 1993, \icarus, 106, 20

\bibitem[{{Natta} {et~al.}(2007){Natta}, {Testi}, {Calvet}, {Henning}, {Waters}, \& {Wilner}}]{Natta2007}
{Natta}, A., {Testi}, L., {Calvet}, N., {et~al.} 2007, in Protostars and Planets V, ed. B.~{Reipurth}, D.~{Jewitt}, \& K.~{Keil}, 767

\bibitem[{{Painter} {et~al.}(2025){Painter}, {Andrews}, {Chandler}, {Ueda}, {Wilner}, {Long}, {Macias}, {Carrasco-Gonzalez}, {Chung}, {Liu}, {Birnstiel}, \& {Hughes}}]{Painter2025}
{Painter}, C., {Andrews}, S.~M., {Chandler}, C.~J., {et~al.} 2025, The Open Journal of Astrophysics, 8, 134

\bibitem[{{Pecaut} \& {Mamajek}(2016)}]{Pecaut2016}
{Pecaut}, M.~J., \& {Mamajek}, E.~E. 2016, \mnras, 461, 794

\bibitem[{{P{\'e}rez} {et~al.}(2012){P{\'e}rez}, {Carpenter}, {Chandler}, {Isella}, {Andrews}, {Ricci}, {Calvet}, {Corder}, {Deller}, {Dullemond}, {Greaves}, {Harris}, {Henning}, {Kwon}, {Lazio}, {Linz}, {Mundy}, {Sargent}, {Storm}, {Testi}, \& {Wilner}}]{Perez2012}
{P{\'e}rez}, L.~M., {Carpenter}, J.~M., {Chandler}, C.~J., {et~al.} 2012, \apjl, 760, L17

\bibitem[{{Posch} {et~al.}(2024){Posch}, {Alves}, {Mir{\'e}t-Roig}, {Ratzenb{\"o}ck}, {Gro{\ss}schedl}, {Meingast}, {Swiggum}, \& {Konietzka}}]{Posch2024}
{Posch}, L., {Alves}, J., {Mir{\'e}t-Roig}, N., {et~al.} 2024, arXiv e-prints, arXiv:2410.18080

\bibitem[{{Ricci} {et~al.}(2010{\natexlab{a}}){Ricci}, {Testi}, {Natta}, \& {Brooks}}]{Ricci2010Ophiuchus}
{Ricci}, L., {Testi}, L., {Natta}, A., \& {Brooks}, K.~J. 2010{\natexlab{a}}, \aap, 521, A66

\bibitem[{{Ricci} {et~al.}(2010{\natexlab{b}}){Ricci}, {Testi}, {Natta}, {Neri}, {Cabrit}, \& {Herczeg}}]{Ricci2010Taurus}
{Ricci}, L., {Testi}, L., {Natta}, A., {et~al.} 2010{\natexlab{b}}, \aap, 512, A15

\bibitem[{{Ricci} {et~al.}(2012){Ricci}, {Trotta}, {Testi}, {Natta}, {Isella}, \& {Wilner}}]{Ricci2012}
{Ricci}, L., {Trotta}, F., {Testi}, L., {et~al.} 2012, \aap, 540, A6

\bibitem[{{Rota} {et~al.}(2024){Rota}, {Meijerhof}, {van der Marel}, {Francis}, {van der Tak}, \& {Sellek}}]{Rota2024}
{Rota}, A.~A., {Meijerhof}, J.~D., {van der Marel}, N., {et~al.} 2024, \aap, 684, A134

\bibitem[{{Tazzari} {et~al.}(2021){Tazzari}, {Testi}, {Natta}, {Williams}, {Ansdell}, {Carpenter}, {Facchini}, {Guidi}, {Hogherheijde}, {Manara}, {Miotello}, \& {van der Marel}}]{Tazzari2021}
{Tazzari}, M., {Testi}, L., {Natta}, A., {et~al.} 2021, \mnras, 506, 5117

\bibitem[{{Testi} {et~al.}(2022){Testi}, {Natta}, {Manara}, {de Gregorio Monsalvo}, {Lodato}, {Lopez}, {Muzic}, {Pascucci}, {Sanchis}, {Miranda}, {Scholz}, {De Simone}, \& {Williams}}]{Testi2022}
{Testi}, L., {Natta}, A., {Manara}, C.~F., {et~al.} 2022, \aap, 663, A98

\bibitem[{{Tobin} {et~al.}(2020){Tobin}, {Sheehan}, {Megeath}, {D{\'\i}az-Rodr{\'\i}guez}, {Offner}, {Murillo}, {van 't Hoff}, {van Dishoeck}, {Osorio}, {Anglada}, {Furlan}, {Stutz}, {Reynolds}, {Karnath}, {Fischer}, {Persson}, {Looney}, {Li}, {Stephens}, {Chandler}, {Cox}, {Dunham}, {Tychoniec}, {Kama}, {Kratter}, {Kounkel}, {Mazur}, {Maud}, {Patel}, {Perez}, {Sadavoy}, {Segura-Cox}, {Sharma}, {Stephenson}, {Watson}, \& {Wyrowski}}]{Tobin2020}
{Tobin}, J.~J., {Sheehan}, P.~D., {Megeath}, S.~T., {et~al.} 2020, \apj, 890, 130

\bibitem[{{Tsukagoshi} {et~al.}(2016){Tsukagoshi}, {Nomura}, {Muto}, {Kawabe}, {Ishimoto}, {Kanagawa}, {Okuzumi}, {Ida}, {Walsh}, \& {Millar}}]{Tsukagoshi2016}
{Tsukagoshi}, T., {Nomura}, H., {Muto}, T., {et~al.} 2016, \apjl, 829, L35

\bibitem[{{Ueda} {et~al.}(2022){Ueda}, {Kataoka}, \& {Tsukagoshi}}]{Ueda2022}
{Ueda}, T., {Kataoka}, A., \& {Tsukagoshi}, T. 2022, \apj, 930, 56

\bibitem[{{White} \& {Hillenbrand}(2004)}]{White2004}
{White}, R.~J., \& {Hillenbrand}, L.~A. 2004, \apj, 616, 998

\end{thebibliography}
\bibliographystyle{apj}

\end{document}